\documentclass[journal]{IEEEtran}

\usepackage{siunitx}
\usepackage{booktabs}
\usepackage{graphics}
\usepackage{float}
\usepackage{siunitx}
\usepackage{booktabs}
\usepackage{hyperref}
\usepackage{soul}
\usepackage{caption}
\usepackage{cite}
\usepackage{amsmath,amssymb,amsfonts}
\usepackage{algorithmic}
\usepackage{graphicx}
\usepackage{textcomp}
\usepackage{seqsplit}
\usepackage{xcolor}
\usepackage{makecell}
\usepackage{enumitem}
\usepackage{babel}
\usepackage{hyperref}

\definecolor{main}{cmyk}{0.05,0.05,0.05,0.85}

\usepackage{hyperref}

\begin{document}

\title{Towards The Creation Of The Future Fish Farm}

\author{
    \IEEEauthorblockN{
    Pavlos Papadopoulos\IEEEauthorrefmark{1}, 
    William J Buchanan\IEEEauthorrefmark{1}, 
    Sarwar Sayeed\IEEEauthorrefmark{1}, 
    Nikolaos Pitropakis\IEEEauthorrefmark{1}}
\IEEEauthorblockA{\IEEEauthorrefmark{1}Blockpass ID Lab, Edinburgh Napier University}
    
  \IEEEauthorblockA{\IEEEauthorrefmark{2} 
}}
\maketitle

\begin{abstract}

\textcolor{black}{\textbf{Aim:} A fish farm is an area where fish raise and bred for food. Fish farm environments support the care and management of seafood within a controlled environment. Over the past few decades, there has been a remarkable increase in the calorie intake of protein attributed to seafood. Along with this, there are significant opportunities within the fish farming industry for economic development. Determining the fish diseases, monitoring the aquatic organisms, and examining the imbalance in the water element are some key factors that require precise observation to determine the accuracy of the acquired data. Similarly, due to the rapid expansion of aquaculture, new technologies are constantly being implemented in this sector to enhance efficiency. However, the existing approaches have often failed to provide an efficient method of farming fish.}

\textcolor{black}{\textbf{Methods:} This work has kept aside the traditional approaches and opened up new dimensions to perform accurate analysis by adopting a distributed ledger technology. Our work analyses the current state-of-the-art of fish farming and proposes a fish farm ecosystem that relies on a private-by-design architecture based on the Hyperledger Fabric private-permissioned distributed ledger technology.} 

\textcolor{black}{\textbf{Results:} The proposed method puts forward accurate and secure storage of the retrieved data from multiple sensors across the ecosystem so that the adhering entities can exercise their decision based on the acquired data.} 

\textcolor{black}{\textbf{Conclusion:} This study demonstrates a proof-of-concept to signify the efficiency and usability of the future fish farm.}

\textcolor{black}{\textbf{Keywords:} blockchain, hyperledger fabric, fish farm, security, privacy, trust}

\end{abstract}

\section{Introduction}

\textcolor{black}{The aquaculture concept is a farming approach that comprises a similar method as agriculture but involves farming aquatic organisms such as fish rather than plants~\cite{why_fish_farm_necessary}. Farming fish not only helps reduce the seafood sully gap but also provides a way to acquire an environmentally friendly protein option. Moreover, compared to other protein resources, it is also an efficient option for consumers. Aquaculture can comprise either extensive or intensive production approaches~\cite{Aqua_culture}. Extensive aquaculture can have very little monitoring over the environment of the cultured organism, whereas intensive aquaculture is based on a highly controlled environment, which may include monitoring several requirements such as temperature, dissolved oxygen, and diet conserved within particular desired levels.}

\textcolor{black}{A fish farm, which is a water-based agriculture, is a subset of aquaculture. Fish farming is increasing rapidly in order to sustain the growth of fish as a protein source~\cite{Fish_farm_definition}.} About 62.5\% of the world's farmed fish are produced by utilising rivers, lakes, and fish farms, whereas the core functionalities of a fish farm can include breeding and hatching fish. A fish farm can use fresh water, sea, salt water, or brackish water to perform its operation. 
There are various factors that aquaculture needs to ensure when farming fish. Food is an essential substance as it supplies energy inputs to contain proper growth~\cite{Aqua_culture}. Likewise, the demand for feed constantly changes in the fish farm ecosystem as the species continue to evolve. However, a traditional fish farm fails to generate continuous allocation of food, thus resulting in vast numbers of mortality. \textcolor{black}{Moreover, water is also a crucial element in a fish farm and the key parameter required for the survival of major species. However, it may not always be possible to maintain the water quality variables at proper levels, in order to ensure maximal growth. To tackle those challenges, collecting accurate data from multiple different sensors across the fish farm ecosystem is very important.}

Overall, the demand for seafood continues to increase, and seafood consumption has doubled over the past five decades~\cite{hang2020secure}. On top of that, around 15\% of the protein-calorie intake worldwide is related to seafood. 

The seafood industry can also support economic development within rural areas. In Scotland, for example, the Scottish Government has defined aquaculture as a critical area of economic development~\cite{aqua}. This includes areas around fish farming, especially in the north and west of Scotland. The key objective is \emph{supporting a healthy and sustainable Scottish aquaculture industry through world-leading science and research} \cite{aqua}. 

\textcolor{black}{While many fish farms provide local data gathering capabilities, sharing the gathered data is often not supported. Additionally, the remote nature of farms makes gathering data difficult due to the expense involved in setting up remote communication channels. Satellite-gathered data fed directly into a cloud environment through satellite communications can thus offer many benefits to localised data gathering. However, the privacy of this type of communication is challenging and often questioned \cite{mckenna2018role,stoyanova2020survey}. Additionally, the security and privacy of the collected data is an ongoing challenge that can only be assured via fundamentally secure digital technologies and approaches \cite{amiri2022big}.} 

\textcolor{black}{The rapid adoption of blockchain has transformed the operations of aquaculture, resolving many insoluble challenges, whereas, at the same time, it helps store trusted data in an immutable way while accelerating the overall processing of the endorsed task. Our work thus outlines the creation of a private-permissioned blockchain infrastructure for the collection of data from multiple sensors within a fish farm environment. While many fish farms provide local data gathering, there is often a lack of sharing of the gathered data and multiple security and privacy concerns \cite{andreas2021towards}. The remote nature of farms often makes gathering data difficult due to the introduced expenses involved in setting up remote communication channels. Our work manages to thoroughly investigate the state-of-the-art approaches, finally proposing a modern blockchain-accelerated connected fish farm system within Scotland.}

\subsection{Fish Farming And Modern Approaches}
Wang et al. ~\cite{wang2020current}, after having monitored and analysed fish farming in China, came to the conclusion that the growing complexity of integrated fish farming required increased attention from the scientific community. Choi et al. ~\cite{choi2020study} suggested that the increasing demand for fishery products, along with the identified limitations within the fishing industry, could be potentially addressed by the aquaculture industry by providing fish stocks. Sangirova et al. ~\cite{sangirova2020benefits} also supported that fish farming can reduce the cost of fish while maintaining the supply of many types of commercial fish. By 2030, it is projected that aquaculture will account for 60\% of the production and 40\% from fishing \cite{cordova2019cloud}.

The key elements of maintaining the health of the fish within a fish farm relate directly to the quality of the water environment provided \cite{arafat2020dataset}, and can be seen as follows:
\begin{itemize}
\item \emph{Turbidity level}. \textcolor{black}{Turbidity measures the cloudiness or haziness of a fluid and is using the units of Nephelometric Turbidity Units (NTU). If there is a significant concentration of suspended material in the water, it will appear as dirty. High levels of algae can create this issue, resulting in harming fish, such as in the case of Trichodiniasis. High turbidity levels can also affect the proper growth of fish eggs and larvae \cite{auld1978effects} by introducing levels of poisoning.} 
\item \emph{pH level}. Different types of fish prefer different pH conditions. The levels between 9–14 can damage the cellular membranes of a fish, while low pH levels can cause rock material in the sediment to relate metals into the water (and thus increase turbidity). 
\item \emph{Temperature level}. \textcolor{black}{Most freshwater fish are cold-blooded and absorb warmth from their surroundings. Thus, it affects their metabolism, and rapid temperature changes can reduce their growth and cause stress to the fish \cite{el1996effects}.}
\item \emph{Dissolved oxygen (DO)}. Multiple studies have shown that the dissolved oxygen levels in water can significantly affect the well-being of fish \cite{solstorm2018dissolved}\cite{null2017dissolved}. DO is measured in mg/L. 
\end{itemize}

In 2015, Chen et al. \cite{chen2015automated} defined an automated environment for fish farming consisted of 
a number of different sensors, such as temperature sensors, dissolved oxygen, pH sensors and water level sensors to monitor fish within a tank. They also used ultrasound to determine the water levels in the tanks. In terms of outputs, the main actuators were: i) RGB light modulation system, to control light outputs by driving different colours of light and different intensities, ii)  Heaters to heat the water to the required temperature, iii) Inflators to add oxygen into the tanks whenever the dissolved oxygen value falls below a given value, iv) Feeders to feed fish at any given times and v) Power supplies to support the sensor infrastructure and act as a fail-safe in case a power issue emerges.
Kim et al. \cite{kim2018realization} implemented a fish farm infrastructure using a range of sensors and actuators. 
Within their system they created a private network 
with sensors connected to an Oxyguard unit and an Arduino. 




Ullah et al. \cite{ullah2018optimization} developed a method to optimize the water pump control, thus maintaining the desired water level by efficiently consuming energy. 
\textcolor{black}{This is related to the pump flow rate and the tank filling level, using Message Queue Telemetry Transport (MQTT) for the control loops while applying a Kalman filter to remove sensor errors. Taniguchi et al. \cite{taniguchi2015evaluation} also used ultrasound to monitor fish movements, while Angani et al. \cite{angani2019realization} used Artificial Intelligence (AI) within an Eel Fish Farm, along with an IoT infrastructure and MQTT. Lee et al. \cite{lee2020realization} defined a method to optimize the water process control for water recirculating.}



\textcolor{black}{Quek \cite{quek2020implementation} identified a need for resilience of power supplies within offshore fish farms, proposing the implementation of an IoT-based Direct Current (DC) nanogrid, which used photovoltaic panels.}
Arafat et al. \cite{arafat2020dataset} defined a data set of IoT-related fish farm data focusing on monitoring the water quality. Their dataset contains 9,623 data records, including temperature, pH factor and turbidity data for two different water levels. 

Yang et al. \cite{yang2021deep} outlined methods for applying deep learning, including live fish identification, species classification, behavioural analysis, feeding decisions, size or biomass estimation, and water quality prediction. 
For fish identification, Yang et al. identified the usage of the Fish4Knowledge (F4K) \cite{salman2016fish,sun2018transferring,fisher2016fish4knowledge} and Croatian fish data sets \cite{zhao2018semi}. Another common data set is LifeCLEF 2015 (LCF-15) \cite{joly2015lifeclef} which is extracted from F4K with 93 underwater videos with 15 fish species. It contains class labels with 20,000 sample images. The two most popular machine learning methods for fish identification are Convolutional Neural Network (CNN) and Region-based CNN (R-CNN) \cite{labao2019cascaded}, with CNN being 15\% and 10\% more accurate than SVM and Softmax, respectively. Meng et al. \cite{meng2018underwater} used images of fish captured from Google to train the CNN, while Naddaf et al. \cite{naddaf2018design} used video recordings from Remotely Operated Vehicles (ROV).
Salman et al. \cite{salman2020automatic} used TensorFlow for CNN using theF4K and LCF-15 data sets \cite{joly2015lifeclef}. 

One of the issues with CNN approaches is that they need to be trained through supervised learning, and the quality of the model produced depends on the quality of the training sets. For that reason, the modified deep convolutional Generative Adversarial Network approach of Zhao \cite{zhao2018semi} used a semi-supervised Deep Learning (DL) model. \textcolor{black}{To overcome the difficulty in accessing training data, Mahmood et al. \cite{mahmood2020automatic} used synthetic data and an object detector approach, and created the You Only Look Once (YOLO) v3 method.} 

There are more than 33,000 different species of fish \cite{oosting2019unlocking, yang2021deep}, which vary in size, shape and colour. 
\textcolor{black}{Unfortunately, there can be many environmental changes and variations which may distort the classification. A deep learning model will often try to learn about these changes and make compensations. Again CNN methods are most often used for this. Siddiqui et al.  \cite{siddiqui2018automatic} used CNN and achieved a success rate of 94.3\%, while Salman et al \cite{salman2016fish} achieved an accuracy of over 90\% and compared CNN against other methods such as SVM, KNN, SRC, PCA-SVM, PCA-KNN, CNNSVM, and CNN-KNN for the LifeCLEF14 \cite{glotinlifeclef} and LifeCLEF15 \cite{joly2015lifeclef} data sets. Along with visual methods, sound has also been used to identify species, such as when Ibrahim et al. \cite{ibrahim2018automatic} used CNN and Long Short-Term Memory (LSTM) models and achieved an accuracy of around 90\%.}


Along with fish identification and classification, the care of fish often requires monitoring their behaviour, especially to support capturing and feeding decisions \cite{papadakis2012computer}. Deep learning has thus been used based on time-series analysis and the ability to recognise visual patterns. CNN \cite{li2019detection, romero2019idtracker, yang2021deep} and Recurrent Neural Network (RNN) methods have been applied as they are useful in detecting localised behaviours \cite{zhao2018modified}. This has included crossing, overlapping and blocking the detection of fish populations. 

A key element of effective planning in fish farms is the abundance, quantity, size and weight of the managed fish population \cite{yang2021deep}. This is often estimated using length, width, weight and area characteristics. However, it can be challenging to monitor due to environmental conditions (such as variations in light intensity and water visibility), thus making necessary the application of methods using CNN \cite{levy2018automated}, R-CNN \cite{alvarez2020image} and Generative Adversarial Network (GAN) \cite{levy2018automated}.

An important element within breeding and production efficiency is the feeding level given to the fish, which can be one of the most costly elements in the fish farming environment.
There are many factors related to feeding, including physiological, nutritional, environmental, and husbandry factors \cite{sun2016models}. 
M{\aa}l{\o}y et al. \cite{maaloy2019spatio, yang2021deep} used temporal and spatial flow with three-dimensional CNN (3D-CNN) and RNN to recognise feeding and non-feeding behaviours. 

As previously mentioned, water quality is a key factor within the environment for fish production, and where dissolved oxygen provides one of the most important factors. Unfortunately, there can be a lag in the supply of oxygen and its effect on water quality. DL methods address this and create a prediction model by using CNN/LSTM \cite{ta2018research}, RNN \cite{liu2019attention} and a Deep Belief Network (DBN) \cite{lin2018deep}. 
Cordova-Rozas al. \cite{cordova2019cloud} focused on water quality for their cloud-based monitoring system. 
Their system monitored fish species in an aquarium of $3m \times 1m \times 2m$ in Peru.
Siva Kumar et al. \cite{sivakumar2021intuitive} also focused on water quality for their cloud-based system for smart aquaculture, and monitored temperature, pH, DO, and Ammonia by using the Blynk private cloud integrated framework \cite{blynk} to collect data in real-time. 

\textcolor{black}{Tawfeeq et al. \cite{tawfeeq2019iot} also implemented a cloud-based infrastructure for a fish farm in Omar, by integrating it into a Wi-Fi network with ESP8266 and a cloud database of Things Speak \cite{things} to gather temperature, water level, pH and Turbidity.} 
Dzulqornain et al. \cite{dzulqornain2018design} outlined an aquaculture based on the "If This Then That" (IFTTT) model and cloud integration. The smart sensors included dissolved oxygen, the potential of hydrogen, water temperature and water level 
within a pond area of $4m \times 5m$. 


\subsection{Related Work}
\label{sec:Related_Works}
Hang et al. \cite{hang2020secure} defined a secure fish farm platform which uses blockchain to achieve trust.
According to their solution, a smart contract is used to automate data gathering, and Hyperledger Fabric is used to create a prototype. Their system included a fish farm contract and a policy contract.
The data gathered for the fish farm contract included: outlier filtering, water level, temperature level, and oxygen level, which then controlled a water pump. For the policy contract, the entities involved included a farmer, a farm owner, multiple devices, the network access policy and a business access policy. 
Regarding the trust, each entity - including the farmer, the farm owner and each device has a public and private key.
These keys are used to identify the identities of each entity and are issued to a certificate authority. A revocation request is then issued if there has been a breach of the entity's private key. Elements of the transactions are: Collect Water Level, Predicted Water Level, Energy Consumption, Control Water Pump, User Management Farm, Sensor Management Farm, Actuator Management, Predicted Water Level History, Energy Consumption History, and Water Pump History.

\textcolor{black}{To the best of our knowledge, our work is the first to introduce a fish farm developed with the privacy by design principle. Compared to the existing literature solutions, it allows specific participating organisations to query sensitive stored data according to their identity credentials, whereas it also blocks access from other non-verified participants. In specific, by enabling the use of a privacy-preserving feature, we thus allow fish farms to store sensitive data related to their business continuity strategies while eliminating the risk of getting compromised by their sensor providers. We should not neglect the fact that third-party providers are responsible for the economic decay of a variety of different organisations due to their inefficient security controls \cite{dubendorfer2004economic}. It should be highlighted that in the related literature, during an insider attack scenario, a compromised sensor provider could be able to exfiltrate sensitive data collected by the sensors provided to the fish farm. The collected data can later be sold to the highest bidder, thus increasing even more the profit for the malicious parties behind the attacks. Such an attack is not feasible in the scope of our solution.}


\textcolor{black}{We can summarise the main contributions of our work, as follows:}

\begin{itemize}

\item \textcolor{black}{We propose a novel distributed fish farm approach, the first of its kind to introduce the privacy by design feature while maintaining its coherence and robustness.}

\item \textcolor{black}{We implement our suggested solution by leveraging Hyperledger Fabric's private data collection feature, thus creating a secure and private by design smart fish farm.}

\item \textcolor{black}{We establish criteria based on the known literature and then empirically evaluate both the performance and robustness of our smart fish farm.}




\end{itemize}

\textcolor{black}{This paper is organised as follows; 
Section~\ref{sec:Background} details the methods and architecture used for the proposed implementation. Additionally, it explains the permissioned blockchain technology by focusing on the overall functionalities and policies of Hyperledger Fabric.
Section~\ref{sec:implementationresults} firstly presents the specifics about the implementation of the proposed future fish farm, and secondly, thoroughly presents the results and experimentally evaluates the metrics to evidence its efficiency and security. 
Finally, Section~\ref{sec:conclusion} discusses and draws the conclusions while offering some pointers for future work.}

\section{Methods}
\label{sec:Background}


\subsection{Distributed Ledger Technologies and Hyperledger Fabric}

A Distributed Ledger Technology (DLT) refers to the database which remains synchronised across many different locations~\cite{difference_DLT_Blockchain}. Its decentralised nature eliminates the necessity of an intermediary in order to validate or authenticate transactions. Blockchain is one of the most important innovations of recent years, bringing vast advancements by transforming traditional centralised approaches. It has appeared as a game-changer in the technology field and is currently being implemented in almost every sector. \textcolor{black}{Blockchain comprises public and private options, where the validators and end-users can be given access based on the platform they joined. As our suggested future fish farm architecture is implemented utilising a private-permissioned blockchain, we emphasise on core functionalities of such approaches.}  

\textcolor{black}{Hyperledger Fabric is a project supported by the Linux foundation. It is designed to form a private-permissioned blockchain architecture which can be leveraged in a multi-organisational approach where each organisation is connected to each other.} Several key aspects make Hyperledger Fabric distinctive and robust compared to other approaches: 
\begin{itemize}

\item Privacy: \textcolor{black}{Hyperledger Fabric requires all of the nodes within a channel to be identified via a Membership Service Provider (MSP). The process is referred to as "private" membership as unlike public blockchains, such as bitcoin, only authorised members are permitted to join the Hyperledger Fabric network.} Hyperledger Fabric is an eminent option for many enterprises and farms concerned about their data privacy. 
Furthermore, Hyperledger Fabric provides flexible design options for the architecture according to the requirements; hence, the necessity for the permissions can be flexible and set according to the requirements.  

\item Channels: Hyperledger Fabric comprises this unique feature which enables it to partition the blockchain ledger into separate channels, thus allowing the peer nodes to generate a separate set of transactions which can be isolated from other parts of the network. \textcolor{black}{This approach is efficient when the architecture is formed with several domains and sensitive data required to be segregated from other entities within the network.} 

\item Scalability: \textcolor{black}{Scalability is another notable characteristic of Hyperleder Fabric, especially when creating a large-scale architecture, since, regardless of the number of nodes, the participating nodes can scale quickly, whereas the system is still able to execute significant amounts of data with minimal resources. This is very helpful when a blockchain infrastructure is developed with a few nodes and the scale is based on demand.} 

\item Modularity: \textcolor{black}{Modularity is another advantage which makes Hyperledger Fabric unique from other blockchain platforms. Hyperledger Fabric is designed to allow separate components to be added and implemented at various stages. Moreover, many components are optional; therefore, those can be removed entirely or initiated at a later stage if required. This offers the authority to the associated domains to determine what parts are necessary to implement at what stage. Some of the modular or "plug-and-play" components that Hyperledger Fabric comprises are consensus, ledger storage, particular access to APIs, and integration of chaincode.}

\end{itemize}


In Hyperledger Fabric, depending on the acquired policies, all transactions are required to be validated by the majority of the nodes within the network~\cite{gorenflo2020fastfabric}. The whole process of transaction validation occurs in a few stages, which is often referred to as consensus. The process of validating, committing and approving the chaincode occurs through a consensus mechanism. Reaching consensus is a process that ensures that the blockchain operates according to the set policies. Hence, the liable nodes are required to provide a guaranteed ordering of the transactions as well as take part in the validation of the transactions.


The complete consensus process in Hyperledger Fabric may consist of 3 phases: Endorsement, Ordering, and Validation. The policy drives the endorsement, requiring endorsing peers to acknowledge it. The ordering nodes set the order that requires to be committed, whereas the validation phase verifies the correctness. Regarding the ordering, in Hyperledger Fabric, some nodes are designated as orderers, ensuring that all the peer nodes comprise the same updated ledger. In a way, the orderer ensures that the consistency of data is maintained to protect the integrity of the blockchain. The peers (or nodes) that are specially designated as orderers ensure that all the peers within a channel have the same updated ledger. In this sense, orderer peers ensure data consistency and protect the ledger's integrity. Orderers also construct the blocks after the endorsement of a transaction and enter the record into them. The orderer peers, collectively known as the ordering service when working in cohesion, mail out the new blocks to each peer within a channel to update their respective ledgers. The ordering service is a modular component. It is important to acknowledge that there are several methods for implementing this ordering service within a Fabric network. Finally, every peer node validates the transactions that are ordered in sequence. Since the transactions are placed by order, the peer nodes can verify if any later transactions were rejected by earlier transactions. Such verification checks will prevent the possibility of double-spending or inconsistency in data.

\textcolor{black}{In Hyperledger Fabric, policies can be defined according to the participating organisations. The endorsement policy specifies that the set of peers on a channel can participate in the transaction validation process by executing chaincode and endorsing the results. Although the endorsement policy does not ensure the correctness of the chaincode on the right peer, another mechanism, “endorsing” and installing chaincode packages, carries out such checks. A few examples of endorsement policies include: i) All peer nodes in the channel can endorse a transaction, ii) A majority of peers in the channel can take part in the endorsement, and iii) At a certain channel, peers must endorse a transaction.}

\subsection{Architecture}
\label{architecture}









\begin{figure*}[h!]
\centering
\includegraphics[width=1\linewidth]{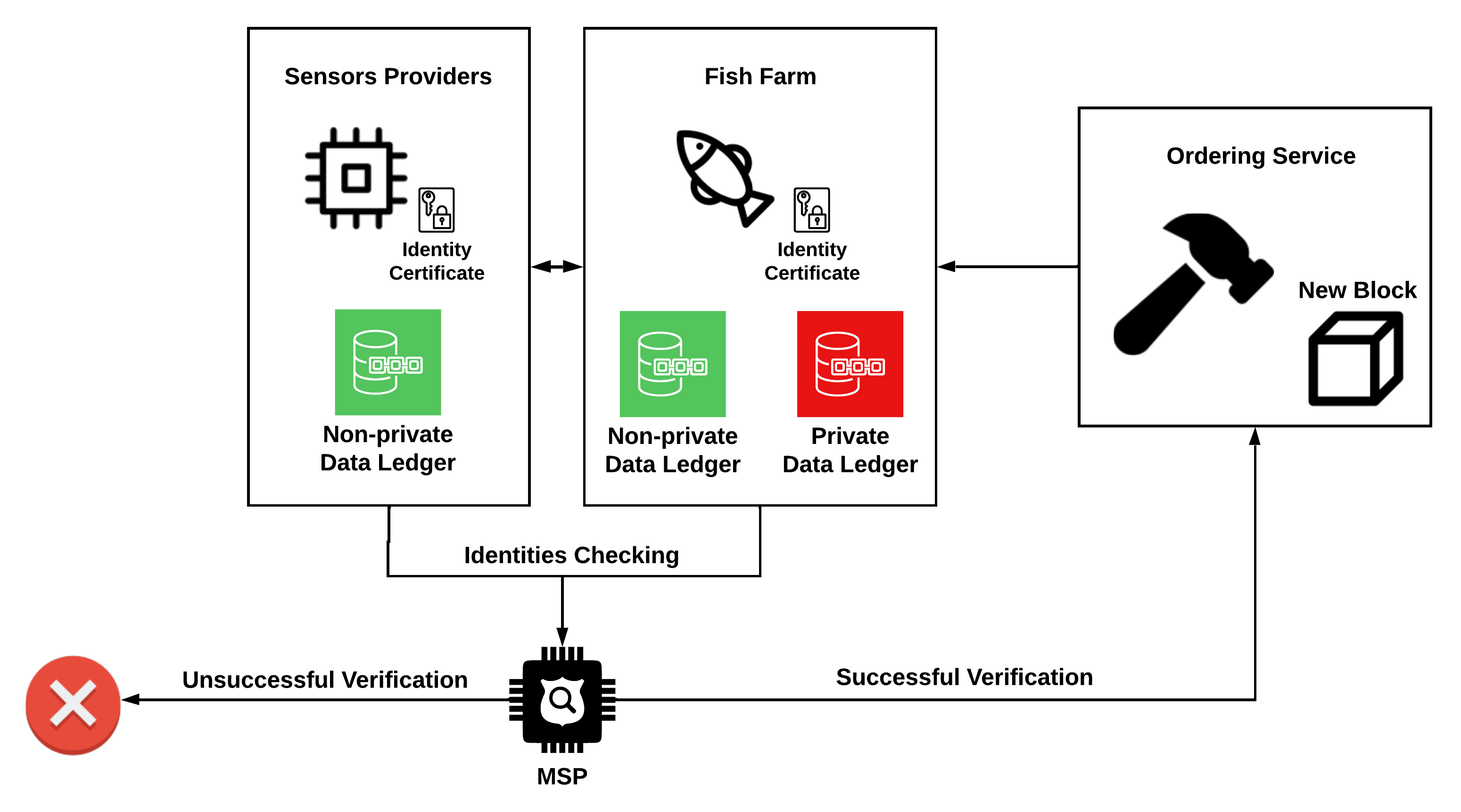}
\caption{Fish Farm architecture overview. The two participating organisations have different levels of access to the system. The Sensors Providers organisation has access only to the non-private data ledger, whereas the Fish Farm organisation also has access to the private data ledger. Access to unauthorised participants is being denied.}
\label{fig:architecture}
\end{figure*}

The architecture of the proposed solution is illustrated in Figure~\ref{fig:architecture}. 
\textcolor{black}{The technical architecture derives from a computational testbed consisting of an Ubuntu 20.04 LTS operating system, with an 8th generation i7 CPU with 6 cores at 3.20GHz, 32GB of RAM, and 1TB SSD. The chosen distributed ledger technology is the Hyperledger Fabric private-permissioned blockchain framework which offers quicker transaction times than other public blockchains \cite{baliga2018performance}. Additionally, since the consensus mechanism in Hyperledger Fabric is flexible, the technology's specifications can be adapted according to the implementation of the use case; hence, the infrastructure can be extended to other similar other use cases that allow the adoption of data-gathering tools \cite{papadopoulos2022decentralized}.}

The topology and the specified technical details of our implementation \cite{papadopoulos2022decentralized} derive as:
\begin{enumerate}
    \item \textcolor{black}{Each sensor provider acts as a Hyperledger Fabric peer with storing access to the blockchain ledger.
    \item Each fish farm is a Hyperledger Fabric peer too, but with viewing access only to the blockchain ledger.} Additionally, the viewing access to the blockchain ledger is further configured to separate each participant's viewing privileges utilising the Private Data Collection feature \cite{stamatellis2020privacy,papadopoulos2020privacy}. This feature is similar to access control policies found in other computational systems.
    \item The peers in our infrastructure, namely \textit{developers.sensorsprovider.org}, \textit{support.sensorsprovider.org}, \textit{admin.fishfarm.org}, and \textit{user.fishfarm.org}, hold the blockchain ledger, the defined private data collections according to the set policies, and record any data tampering. The used state databases that peers are using are CouchDB instances.
    \item The identity of each peer is an X.509 certificate that is being verified by the Membership Service Provider (MSP) entity for its validity.
    \item Group of peers can form Hyperledger Fabric organisations. The role of the organisations in our architecture is to accept/reject each blockchain transaction according to the defined policy. In the technical experimentation, there are two specified organisations, namely \textit{sensorsproviders.org} and \textit{fishfarm.org}.
    \item The ordering service, in our case, the crash-fault tolerant RAFT service, creates the new blockchain blocks and broadcasts them to all the participating peers according to the defined policy. \textcolor{black}{Hence, three orderers handle each storing transaction to avoid potential single point of failures that single-orderer infrastructures face. It should be noted that any number of orderers could be used, and we have specifically chosen three only for experimental purposes. There is no correlation between the number of orderers with the number of other Hyperledger Fabric components.}  
    \item \textcolor{black}{The smart contract of our solution, namely chaincode in Hyperledger Fabric, is being approved and installed in all the peers of the participating organisations and the ordering service. Chaincode is written using the Go programming language.}
    \item \textcolor{black}{We have generated and utilised synthetic data based on the data fields of the infrastructure.}
    \item \textcolor{black}{The infrastructure's policy is specified during the initialisation of the blockchain but can also be further updated to include new blockchain rules.} In update scenarios, the new policy needs to be approved by a number of participating organisations and the ordering service, similar to chaincode updates.
\end{enumerate}

\section{Results}
\label{sec:implementationresults}

\subsection{Proof-of-Concept and Access Control Policy}
This subsection defines the developed Proof-of-Concept (PoC) and our system's detailed access control policy. The PoC involves setting up a permissioned blockchain that is based on Hyperledger Fabric version 2.3.0, using the Minifabric framework\footnote{Minifabric framework: \url{https://github.com/hyperledger-labs/minifabric}}.

There are two distinct private data collections, namely \textit{collectionFishFarm} and \textit{collectionFishFarmPrivateDetails}. It was considered that the \textit{sensorsprovider.org} organisation is the provider of the sensors that monitor the fish farm and the \textit{fishfarm.org} organisation is the fish farm that installs these sensors. However, sensor providers often require access to the sensors for maintenance purposes; consequently, they get access to the data collected from their sensors. In the presented testbed, the fish farm can utilise the infrastructure to reveal only necessary information to the sensors' providers and not expose any collected sensitive details. These sensitive details may include information about specific fish farm metrics that the fish farm can further utilise, commercialise, receive governmental funding and more.

\textcolor{black}{Hence, there are 14 total data fields that derive to data fields that both organisations can access, such as \textit{windspeed}, \textit{rainfall}, \textit{airpressure}, \textit{temperature}, \textit{waveheight}, and \textit{watercurrent}, as well as, private data fields that only the \textit{fishfarm.org} organisation can access, such as \textit{fdom (Fluorescent Dissolved Organic Matter)}, \textit{salinity}, \textit{ph}, \textit{turbidity}, \textit{algae}, \textit{orp (Oxidation-Reduction Potential)}, \textit{nitrates}. Finally, there is an extra data field, namely \textit{name}, which is the \textit{key} that connects the two data collections. As mentioned in our topology, the data utilised in these data fields have been synthetically generated. In a real-world scenario, the equivalent data fields could be collected from physical sensors carefully placed in a fish farm.}

\textcolor{black}{Every node involved in this PoC is developed as a docker container and is authorised by the network administrators prior to joining the channel. The administrators issue digital certificates to the peers in the fish farm ecosystem. The Membership Service Provider (MSP) is responsible for defining the rules by which identities are validated, authenticated, and allowed access to a network~\cite{MSP_CA}. The MSP leverages the Certificate Authority (CA), the entity responsible for creating and revoking identity certificates. Likewise, every entity that is part of the network is issued X.509 certificates. The modular infrastructure of the Hyperledger Fabric permits to impose of external CAs.} 



\subsection{Evaluation}
\label{sec:evaluation}

\begin{figure*}[h!]
\centering
\includegraphics[width=0.8\linewidth]{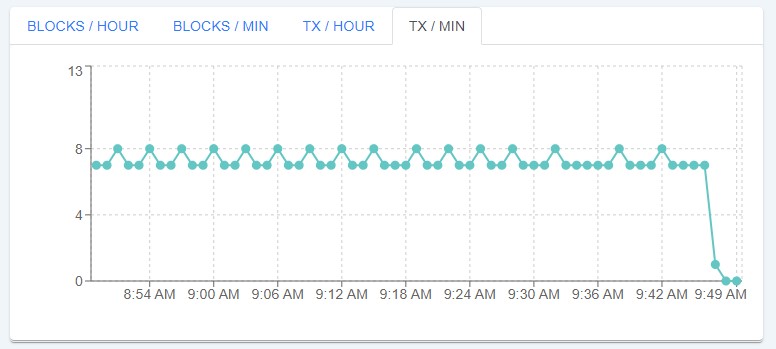}
\caption{\textcolor{black}{Proof-of-Concept's throughput in transactions per minute.}}
\label{fig:txperminute}
\end{figure*}
\begin{figure*}[h!]
\centering
\includegraphics[width=0.8\linewidth]{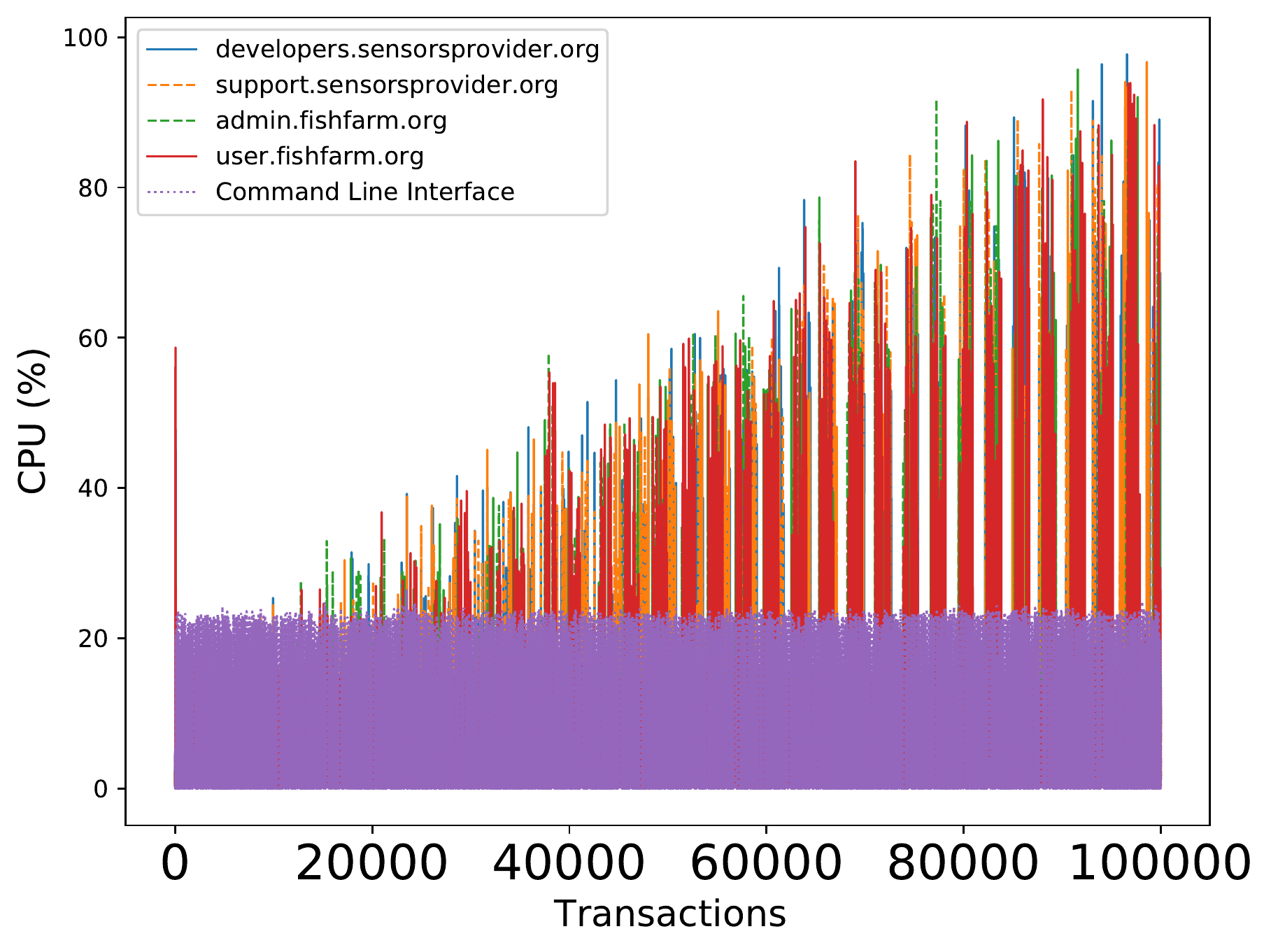}
\caption{CPU usage to write 100,000 transactions to the ledger.}
\label{fig:cpuwrite100k}
\end{figure*}

\textcolor{black}{In this subsection, the system's results and experimental evaluation can be seen in terms of performance and security. As observed in the literature \cite{papadopoulos2020privacy,stamatellis2020privacy}, the Hyperledger Fabric is very efficient compared to other similar systems developed using different technologies. However, despite the fact that the addition of the Minifabric framework aided the PoC's development activities, the system's performance is degraded. A write transaction required approximately 7.3 seconds to be conducted (Appendix~\ref{app:write}), whereas approximately 6.9 seconds were required for a read transaction (Appendix~\ref{app:read}). Hence, as visualised using Hyperledger Explorer, the PoC's throughput is between 7 and 8 transactions per minute, as seen in Figure~\ref{fig:txperminute}. This computational overhead occurs due to the execution of the \textit{minifab} script (part of the Minifabric framework) that manages the Hyperledger Fabric environment. However, this overhead could be avoided in production environments, whereas the infrastructures are being developed using traditional Hyperledger Fabric practices. Although, when the PoC scaled up to 100,000 stored records, the system's performance remained the same, which was an expected outcome of the proposed solution that proves its superiority against other technologies (Appendix~\ref{app:read}). Additionally, the system's performance has been experimentally evaluated and visualised in plots using Python's \textit{Matplotlib}. The CPU performance of our system to store 100,000 records in the system is depicted in Figure~\ref{fig:cpuwrite100k}. This figure shows that the CPU usage of each participating peer constantly fluctuates to store each record in the blockchain system. However, in most cases, these fluctuations occur within the 0-20\% range (as visualised with purple colour regarding the command line interface), with CPU usage spikes for all the participating peers after a certain timeframe. Hence, it is speculated that these CPU usage spikes occur from utilising the Minifabric framework outside of our control, as well as potential hardware limitations and other environmental impacts.}

The CPU performance to Read a record on any number of records, as well as the RAM usage for any read or write transactions, are negligible (< 5\% CPU performance and < 1\% RAM usage); hence, they are not plotted.
Fair performance comparison with other works in the literature that utilise the private data collection feature, such as PREHEALTH \cite{stamatellis2020privacy} and PRESERVE DNS \cite{papadopoulos2020privacy}, cannot be done since they are using the "vanilla" Hyperledger Fabric, instead of Minifabric framework, that the Future Fish Farm is built upon. As reported previously, in this work, a read transaction occurs in approximately 6.9 seconds instead of approximately 0.83 seconds in PREHEALTH and PRESERVE DNS.

Regarding the two data collections, as specified in the previous sections, only the peers of the \textit{fishfarm.org} organisation have access to the private data collection, namely \textit{collectionFishFarmPrivateDetails}, whereas all the peers from the two participating peers have access to the \textit{collectionFishFarm} (Appendix~\ref{app:privatedata}). No other parties can access data stored in these two data collections since their identity certificates are not included in the specified policy (Appendix~\ref{app:privatedata}).

\section{Discussion}
\label{sec:conclusion}

\textcolor{black}{Fish farming is the fastest-thriving channel of animal food production. Half of the fish consumed worldwide is produced within an artificial ambience. In this paper, we have proposed the concept of a future fish farm to demonstrate the intelligent observation of acquired data in order to reach an informed decision.} The architecture of the future fish farm is based on a complex approach where the accuracy and reliability of the acquired data, the decision-making models, and the correlation among various intelligent systems must function correctly. 

\textcolor{black}{Our solution is implemented within our testbed, clearly demonstrating the functionalities that were proposed by our novel architecture, thus resulting in a future fish farm with improved effectiveness and performance efficiency. It should be highlighted that the implemented solution is the first of its kind to enable fish farms to collect sensitive data without risking potential exposure to compromised or malicious sensor providers. In the world of malicious data brokers, a malicious sensor provider may involve the exfiltration of critical fish farm data, thus either selling them to other third parties or even tampering with the data to potentially influence the fish farm to make unnecessary buying decisions. This will have consequences not only for the specific fish farm but for the supply chain as a whole, introducing issues to the business continuity of a variety of organisations. However, our solution disables such attack vectors and guarantees both security and privacy.}


For our future work, we aim to extend the development of the future fish farm infrastructure by adding more functionalities and participants in a more complex scenario that mimics a real-world use case. \textcolor{black}{Additionally, adding AI techniques to gather further insights from the stored data is a compelling future step to determine the usability of the future fish farm in real-world environments.} 


\section*{Declarations}



\subsection{Authors’ contributions}

Made substantial contributions to conception and design of the study and performed data analysis and interpretation: Papadopoulos P., Buchanan W., Sayeed S.,Pitropakis N.

Performed data acquisition, as well as provided administrative, technical, and material support: Papadopoulos P.




\subsection{Availability of data and materials}

Not applicable.



\subsection{Financial support and sponsorship}

None.



\subsection{Conflicts of interest}

All authors declared that there are no conflicts of interest. 





\subsection{Ethical approval and consent to participate}

Not applicable.



\subsection{Consent for publication}
Not applicable.



\subsection{Copyright}

© The Author(s) 2022.

\bibliographystyle{IEEEtran}
\bibliography{template}


\begin{appendices}

\section{Write transaction}
\label{app:write}

\begin{figure*}[h!]
\centering
\includegraphics[width=1\linewidth]{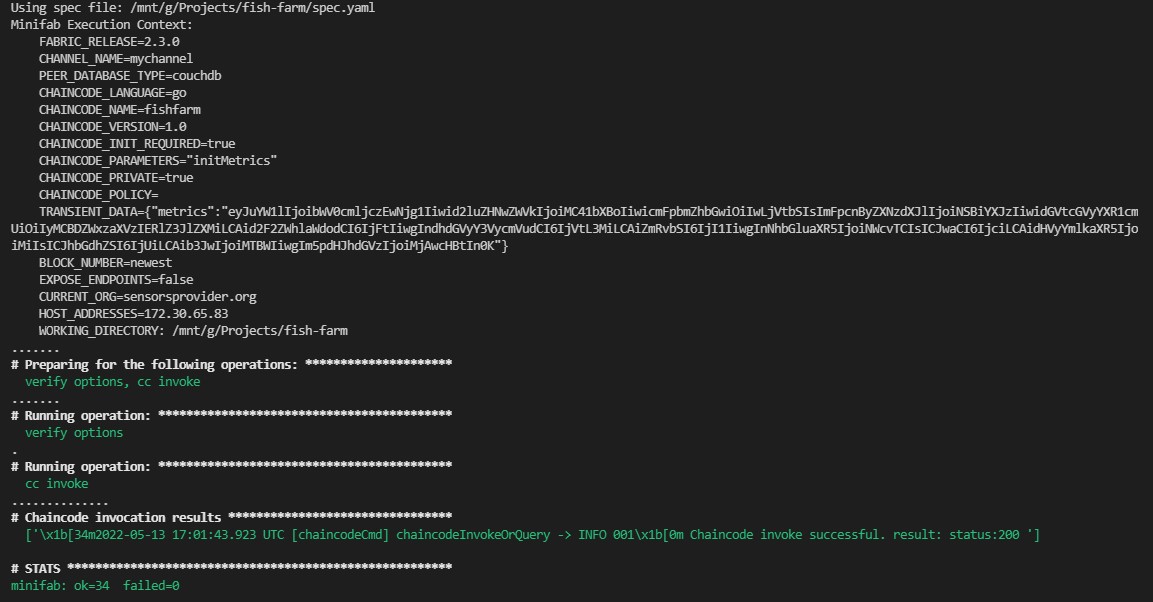}
\caption{Write transaction}
\label{fig:writetransaction}
\end{figure*}

A write transaction can be seen in Figure~\ref{fig:writetransaction}.

\section{Read transaction}
\label{app:read}

\begin{figure*}[h!]
\centering
\includegraphics[width=1\linewidth]{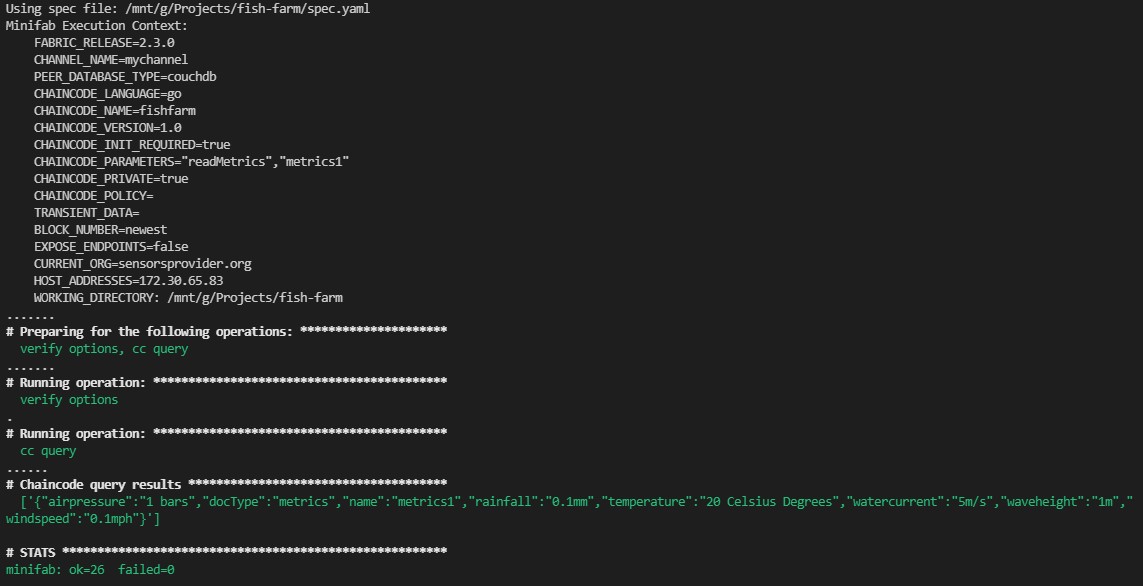}
\caption{Read transaction first record}
\label{fig:readtransaction1}
\end{figure*}

\begin{figure*}[h!]
\centering
\includegraphics[width=1\linewidth]{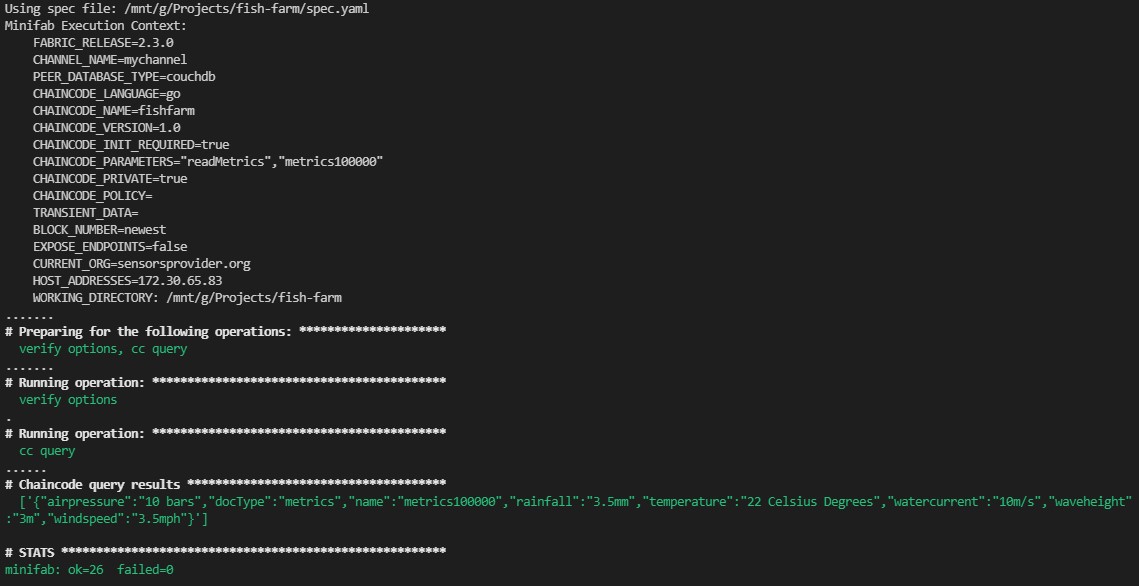}
\caption{Read transaction 100,000th record}
\label{fig:readtransaction100k}
\end{figure*}

The read transaction for the 1st record can be seen in Figure~\ref{fig:readtransaction1}. The read transaction for the 100,000th record can be seen in Figure~\ref{fig:readtransaction100k}.

\section{Private Data Collection}
\label{app:privatedata}

\begin{figure*}[h!]
\centering
\includegraphics[width=1\linewidth]{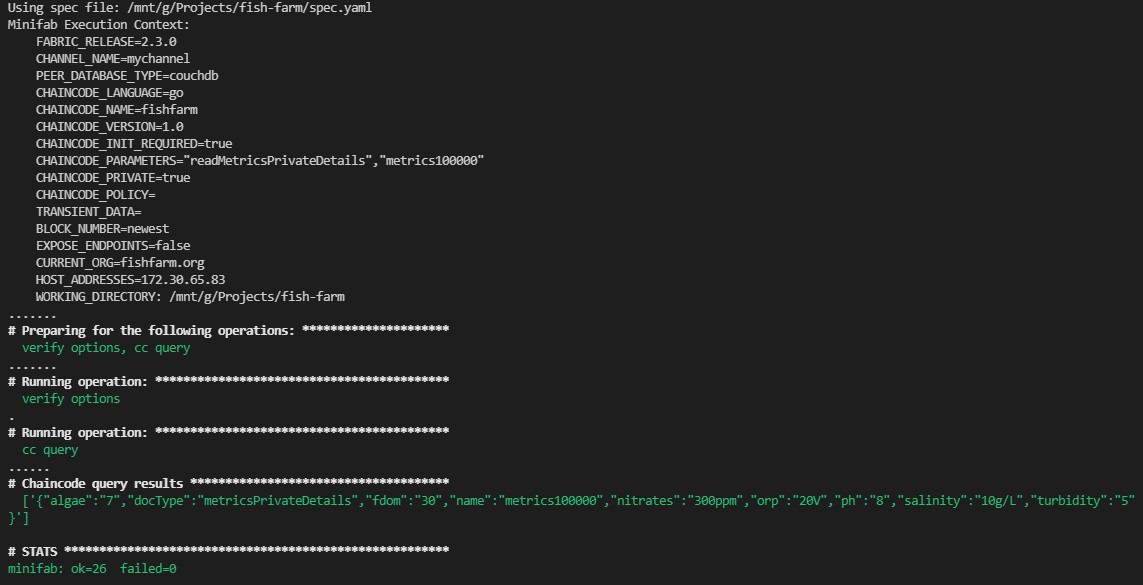}
\caption{Successful read transaction to private data collection}
\label{fig:readprivatesuccess}
\end{figure*}

\begin{figure*}[h!]
\centering
\includegraphics[width=1\linewidth]{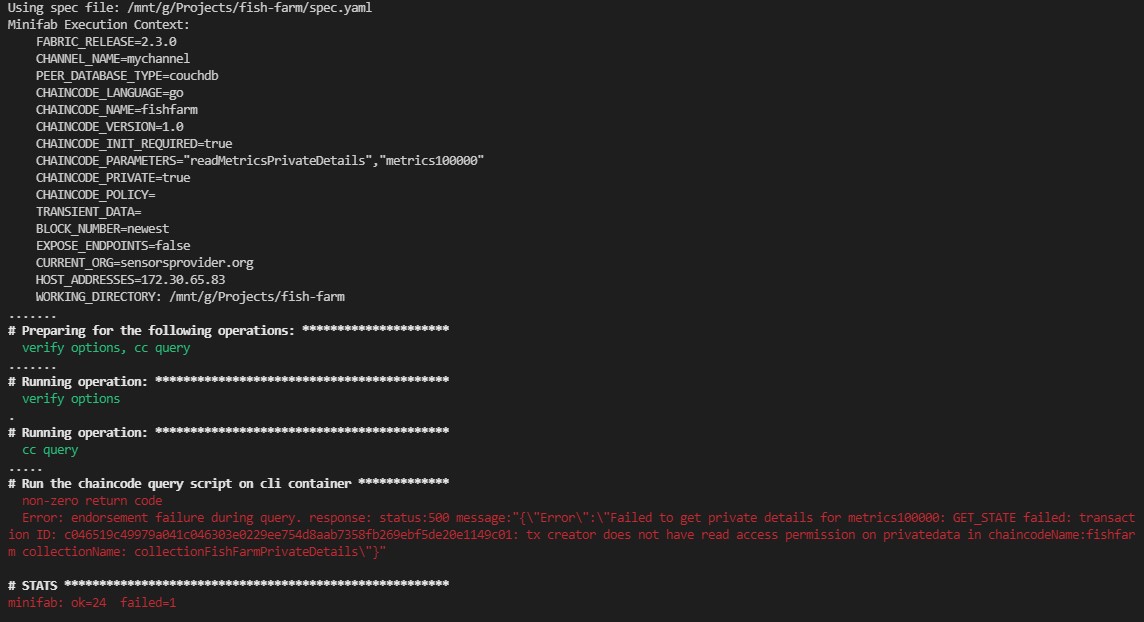}
\caption{Unsuccessful read transaction to private data collection with permission denied error}
\label{fig:readprivatedeny}
\end{figure*}

A successful read transaction to the private data collection can be seen in Figure~\ref{fig:readprivatesuccess}. An unsuccessful read transaction to the private data collection can be seen in Figure~\ref{fig:readprivatedeny}.

\end{appendices}

\end{document}